\documentclass[12pt]{article}
\usepackage{amsmath,amssymb,graphicx,mathrsfs,hyperref,slashed}
\newcommand{\be}{\begin{equation}}
\newcommand{\ee}{\end{equation}}
\newcommand{\bea}{\begin{eqnarray}}
\newcommand{\eea}{\end{eqnarray}}

\def\({\left(} \def\){\right)}

\renewcommand{\baselinestretch}{1.0}
\begin{document}
\title{\vspace{-1.8in}
{Teleporting entanglement  during black hole evaporation}}
\author{\large Ram Brustein${}^{(1)}$,  A.J.M. Medved${}^{(2,3)}$
\\
\vspace{-.5in} \hspace{-1.5in} \vbox{
 \begin{flushleft}
  $^{\textrm{\normalsize
(1)\ Department of Physics, Ben-Gurion University,
    Beer-Sheva 84105, Israel}}$
$^{\textrm{\normalsize (2)\ Department of Physics \& Electronics, Rhodes University,
  Grahamstown 6140, South Africa}}$
$^{\textrm{\normalsize (3)\ National Institute for Theoretical Physics (NITheP), Western Cape 7602,
South Africa}}$
\\ \small \hspace{1.07in}
    ramyb@bgu.ac.il,\  j.medved@ru.ac.za
\end{flushleft}
}}
\date{}
\maketitle
\abstract{
The unitary evaporation of a black hole (BH) in an initially pure state must lead to the eventual purification of the emitted radiation. It follows that the late radiation has to be entangled with the early radiation and, as a consequence, the entanglement among the Hawking pair partners has to decrease continuously from maximal to vanishing during the BH's life span. Starting from the basic premise that both the horizon radius and the center of mass of a finite-mass BH are fluctuating quantum mechanically, we show how this process is realized.  First, it is shown that the horizon fluctuations induce a small amount of variance in the total linear momentum of each created pair. This is in contrast to the case of an infinitely massive BH, for which the total momentum of the produced pair vanishes exactly on account of momentum conservation. This variance leads to a random recoil of the BH  during each emission and, as a result, the center of mass of the BH undergoes a quantum random walk. Consequently, the uncertainty in its momentum grows as the square root of the number of emissions.   We then show that this uncertainty controls the amount of deviation from maximal entanglement of the produced pairs and that this deviation is determined by the ratio of the cumulative number of emitted particles to the initial BH entropy. Thus, the interplay between the horizon and center-of-mass fluctuations provides a mechanism for teleporting entanglement from the pair partners to the BH and the emitted radiation.
}
\newpage
\renewcommand{\baselinestretch}{1.5}\normalsize

\section{Introduction}

Let us consider an evaporating black hole (BH) in a four-dimensional asymptotically flat spacetime. The standard Hawking description of BH radiation
via pair production \cite{Hawk,info} implies that the quantum state of the near-horizon matter is the Unruh vacuum state, which is that
of maximally entangled pairs straddling the horizon. Ultimately, the positive-energy partners fly off to become the Hawking radiation, while their negative-energy counterparts fall into the interior and lead to a reduction in the BH mass.

One problem with this picture is its failure to explain how  information can escape from the BH, which is an essential requirement for a unitary process of evaporation.  It was long thought that this quandary could be resolved by non-perturbative effects such as contributions from other geometries \cite{Mald} or subtle correlations between the emitted quanta \cite{Pagerev}. However, this optimistic stance leads one to an even bigger issue: Explaining how  the information can get out without violating  another fundamental tenet of quantum theory ---  the monogamy of entanglement, {\em i.e.}, that no particle can be strongly entangled with more than one other particle. This concern has  been championed by Mathur \cite{Mathur1,Mathur2,Mathur3,Mathur4,Mathur5}, while  Almheiri {\em et. al.} brought this matter to the forefront with their notion of a   near-horizon  ``firewall''  \cite{AMPS} (also see \cite{Sunny,Hooft,Braun,MP,Bousso}).

Using the strong-subadditivity inequality, one can recast this problem in precise terms, but it is also easy to understand  the central issues at a simple intuitive level.  For information to escape from the BH and then be encoded in the state of the external radiation, there must be some degree of entanglement between the emitted Hawking particles. Otherwise, the final state of the radiation  could have  no ``knowledge'' about the initial state of the collapsing matter. Also, it is not viable for this information to be released  only in the final stages of evaporation, as the amount  that is stored  in the BH interior cannot exceed the horizon area in Planck units. To argue differently would be to argue for BH remnants.  Monogamy of entanglement then  rules out the possibility of maximally entangled pairs in the near-horizon zone --- the Unruh state cannot be the correct quantum state of the near-horizon matter.

One might hope that the entanglement between pair partners could be large enough that the state is still ``approximately Unruh''. However, this hope is squashed once the BH reaches its half life in terms of the number of the emitted particles, the so-called Page time.  The reason being that, according to the Page  model of BH evaporation \cite{page}, which  establishes parametrically the minimal rate of information release for a unitary process,  this is the time when information must begin to emerge from the BH at a rate of  order unity.  This, in turn,  implies that the near-horizon state has significant corrections away from the Unruh state.

The Unruh state had always  been given preferential status in this context because  a freely falling observer in this state would fall  according to the predictions of classical general relativity.  A different choice of vacuum would lead to deviations from these predictions,  and it is implicitly assumed by many that such a choice would be problematic; for example, by  putting the validity of Einstein's equivalence principle at risk. We do not agree, however, that the demise of the equivalence principle follows as a inevitable consequence of having disentangled pairs  \cite{noburn,stick}. In particular, it is shown in \cite{stick}  that, for typical objects,  the classical tidal forces at the  horizon of a finite-mass BH are more dangerous than a significant degree of disentanglement. Our conclusion is  that  one cannot   use this line of reasoning to single out the Unruh vacuum as the preferred state,  and we  will proceed to consider other possibilities in this spirit.

Our basic premise is that a consistent treatment of a finite-mass BH must account for its quantum fluctuations \cite{RB,gia}. The location of the BH horizon is, at least in principle,  a physically measurable quantity \cite{Matt}, so that it makes sense to talk about its quantum fluctuations. Similarly for the position and momentum of the BH center of mass, and so  their quantum spreads also have a physical meaning. So far, we have considered the impact of fluctuations of the horizon area (or radius)  on the BH evaporation process \cite{slowleak,slowburn,flameoff}.  But, as mentioned, the  center of mass of the BH  is also fluctuating quantum mechanically.

 The main new idea of the current paper is that the interplay between the quantum fluctuations of the horizon and those of the center of mass  determine the state of the produced pairs, while  allowing for the swapping  or teleportation of  entanglement from the pair partners to the BH and emitted Hawking radiation.  Our focus is on the entanglement of linear momentum, but we do expect that similar considerations also apply to angular momentum.

In Hawking's model for pair production \cite{info},   the negative-energy partner is  subsumed by the BH interior and  the positive-energy partner escapes to  the exterior. At times much earlier than the Page time, the pair is in a state that is close to  maximally entangled. As the partnership ends with the subsumption of  the negative-energy partner, the state
of the  BH interior will have to change accordingly.  It is unclear what the state of the BH interior is to begin with (however, see \cite{dvali,inny}), never mind what it will change into. But we do know that unitary evolution implies that entanglement will not be destroyed, only teleported. Hence, it must be that the BH interior is now entangled with the positive-energy partner, which is by then part of the exterior radiation.

Ultimately, the BH must teleport  this newly acquired entanglement to its exterior; otherwise, the purification of the external radiation cannot be completed. The only means that the  BH has for doing this is through the  influence of its near-horizon gravitational field on subsequently produced pairs.

So, somehow, the pair-production mechanism needs to ``remember''  the history of emissions! But how? On one hand, the state of the produced pairs is apparently determined solely by the gravitational field near the horizon of the BH. On the other,  a ``memory" requires a dynamical mechanism that allows the history of emissions to change the state of the produced pairs.

Our proposal~\footnote{A similar idea concerning angular momentum was discussed for the case of a rotating BH by Chowdhury and Mathur \cite{MathurXX} in the context of fuzzball models.} is that the  swapping or teleportation of entanglement proceeds via the transfer  of quantum  fluctuations; from those of the horizon  to those of the center of mass.

In the limiting case, when the BH mass is taken to infinity, the momentum transfer vanishes due to momentum conservation in the absence of recoil. For a  BH of finite mass, two new effects arise: The horizon of the BH experiences Planck-sized fluctuations \cite{RM,RB,flucyou} and, additionally, the center of mass of the BH fluctuates. (See Fig.~1.) Our goal is to show  how the interplay between these two types of fluctuations leads  to an increasing uncertainty in knowing one of the partners momentum when the momentum of the other is measured. It is this loss of knowledge that leads  to a decreasing entanglement.

First, we will show that the horizon fluctuations induce a  small variance in the total momentum of each of the created pairs. This variance results in a random recoil of the BH; its center-of-mass momentum  becomes uncertain by a small amount during each emission, and these deviations accumulate. The resulting picture is that the  center-of-mass momentum of the BH undergoes a quantum random walk with  (approximately) fixed step sizes and, hence, the uncertainty in its momentum grows as the square root of the number of emissions.  The growing uncertainty in the center-of-mass momentum leads to an increasing  uncertainty in the correlation between the momenta of the pair partners, which in turn decreases their entanglement. What will eventually be  shown is that this decrease in entanglement is determined by  the ratio of the cumulative number of emitted particles to the initial BH entropy.

\begin{figure}
[t]
\begin{flushleft}
\scalebox{.25}{\includegraphics{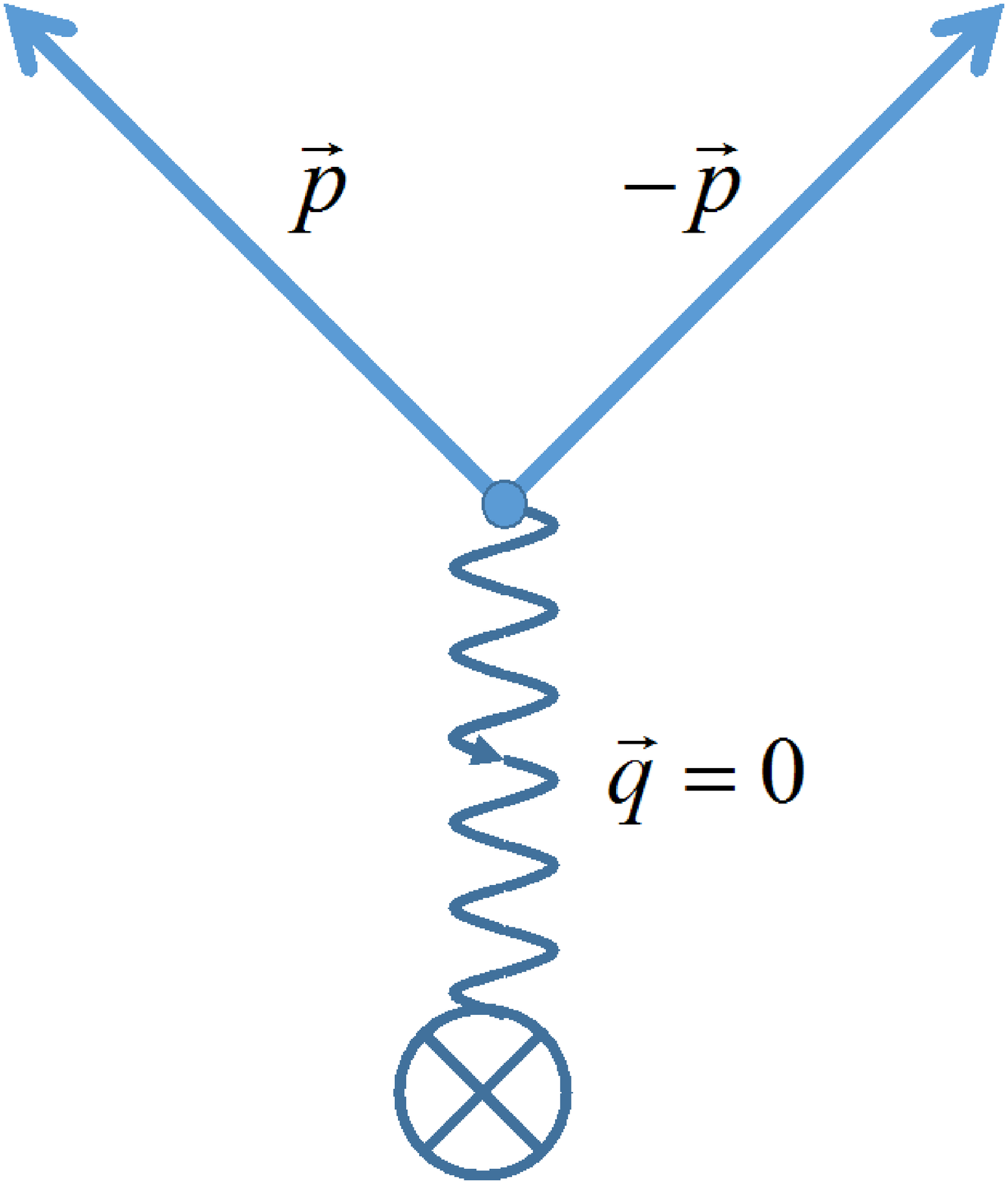}} \scalebox{.25}{\includegraphics{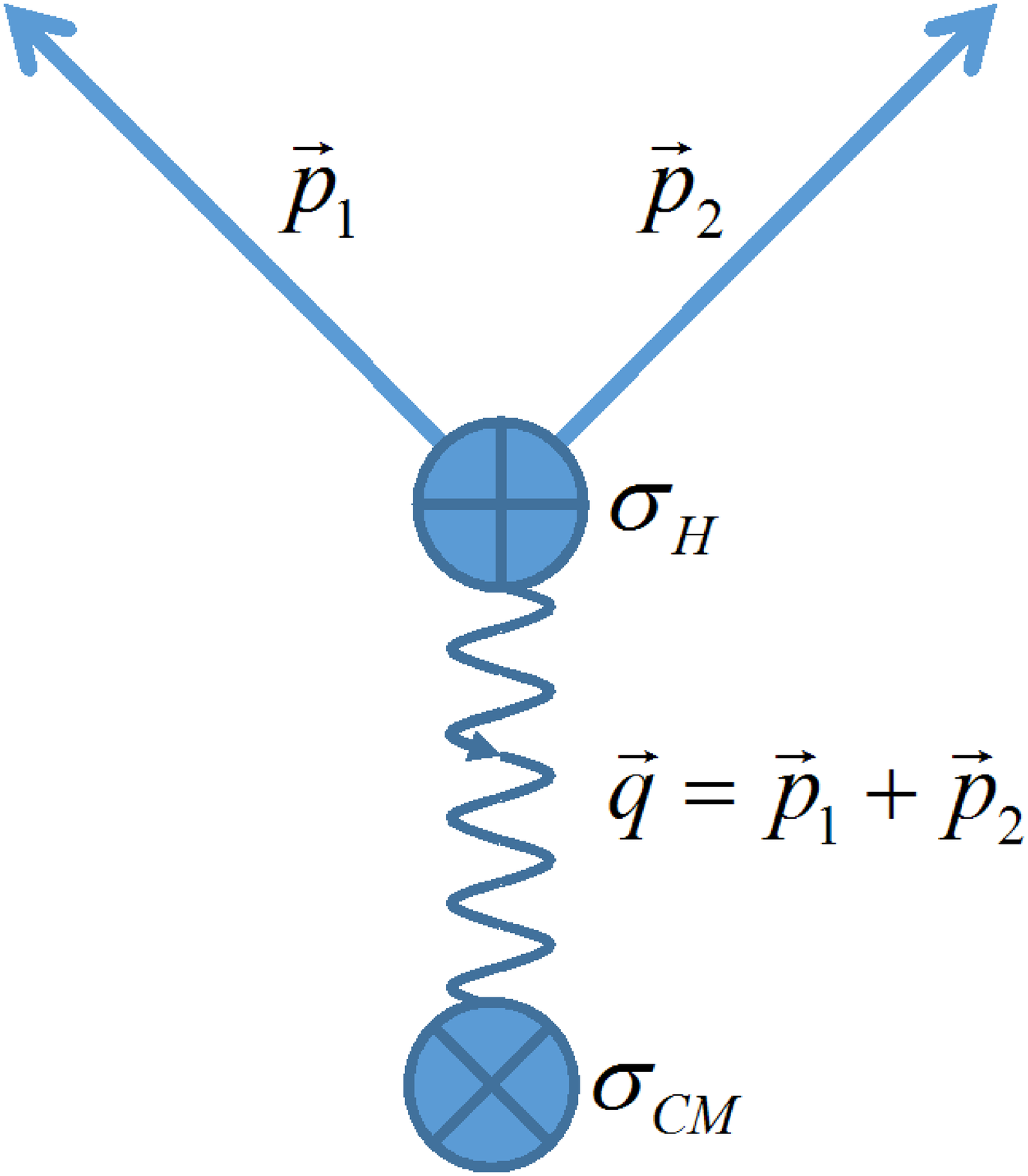}}
\end{flushleft}
\caption{Diagrams of pair production in a BH background. Left: An infinite-mass BH. Neither the horizon position nor center-of-mass (CM) momentum fluctuate, and the momentum transfer vanishes. Right: A finite-mass BH. Both the horizon position and CM momentum fluctuate, and the momentum transfer vanishes on average but not for every pair-production event. }
\end{figure}

There are other sources of momentum fluctuations besides those  induced by the horizon fluctuations. One source is the uncertainty principle. If the initial center-of-mass position of the BH is localized to within a region of size $R_S$, then the center-of-mass momentum has to have a spread of order $1/R_S$ \cite{sunny}. An additional source is the random recoil of the BH due to the eventual absorption of the negative-energy partner of the emitted positive-energy Hawking particles. (See Sect.~6 of \cite{Mathur3}. Also, \cite{Mathurplusplus,Mathurplus}.) This recoil is of order $1/R_S$ per emitted particle. However, since momentum is conserved in this process, it does not lead to any change in the entanglement of the emitted pairs, in agreement with the conclusion of \cite{Mathur3}.

\section{Momentum transfer in  black hole  pair production}

To obtain the  typical scale of  momentum transfer  for a  pair, we first recall that the horizon fluctuates at Planckian scales,~\footnote{Conventions: $R$ is the horizon radius in general, while  $R_S$  is its classical (Schwarzschild) value. Also, $\;T_H\sim \frac{1}{R_S}\;$ is the Hawking temperature and $\;S_{BH}\sim \frac{R_S^2}{l_P^2}\;$ is the Bekenstein--Hawking entropy, where $l_P$ is the Planck length.} $\;\Delta R\sim l_p\;$. Then the  momentum of the partners  fluctuates according to
$
\Delta p\;\sim\; \left.\frac{\partial p}{\partial R}\right|_{R=R_S} \Delta R \;\sim\; \frac{l_p}{R_S^2}\;,
$
where the last estimate is obtained using $\;p\sim T_H \sim 1/R\;$.
This estimate indicates that the typical momentum transfer from the produced pair to the BH is of the extremely small magnitude  $\;q \sim l_p/R_S^2 \;$.

Let us now reinterpret this result in terms of random ``momentum kicks" to the center of mass of the BH. Each pair-production event takes about a Schwarzschild time $R_S$ and results in a  momentum kick of    magnitude $\;\Delta P_{CM}\sim q \sim l_p/R_S^2\;$ in a random direction. This process may be analyzed in terms of a  quantum random walk in momentum space.
In terms of $N$, the cumulative  number of emitted particles, this is
$
\widehat{\vec P}_{CM}(N+1) \;=\; \widehat{\vec P}_{CM}(N)  + \frac{l_p}{R_S^2}
 \widehat{\vec J}\;,
$
where  $\widehat{\vec J}$ is a random Gaussian variable with a mean of zero and a  variance of 1.

The wavefunction of the center-of-mass momentum is therefore a Gaussian with
a  mean of zero and a variance of  $\;\Delta P_{CM}^2 \sim \frac{l_p^2}{R_S^4} N
\sim \frac{N}{S_{BH}} \frac{1}{R_S^2}\;$. Hence,
$
\Psi(P_{CM})\;=\; \frac{1}{\sqrt{{\cal N}}}e^{-\frac{1}{4}\frac{P_{CM}^2}{\sigma_{CM}^2}}\;,
$
with ${\cal N}$ being a suitable normalization constant and
$
\;\sigma_{CM}^2 \;=\;\frac{N}{S_{BH}} \frac{1}{R_S^2} \;.
$

\section{Entanglement in pair production}

What are the entangled quantities in the BH pair-production? They cannot be just the spin or angular momentum because one can imagine a situation when a Schwarzschild BH emits most of the  radiation  in the form of scalar particles.  And just how entangled need they be?

We begin answering these questions by considering the physical observables in the  pair-production process. The factor that  determines the degree of entanglement between the produced particles  is the near-horizon gravitational field. For the case of electron--positron pair production,  this is made clear in
({\em e.g.}) \cite{PP1,PP2}.

Suppose that one measures the momentum of an  outgoing Hawking particle. How accurately would they know about the momentum of its ingoing (negative-energy) partner? If the BH is infinitely massive, the answer would be complete accuracy because momentum is conserved during the pair-production process and, consequently,  the pair would  be maximally entangled. However, what about a BH of finite mass? If the center-of-mass momentum is not measured, the observer would not know what value to assign it and the partner's momentum would  be necessarily uncertain ---  some disentanglement has incurred.  From this point of view, disentanglement results from a lack of information about  one of the pair-member's momentum.

Let us set the initial momentum of the BH center of mass  to zero, $\;\vec{P}_{CM}=0\;$. If the momentum of the first emitted particle is measured, we would know with high precision what the change  to $\vec{P}_{CM}$ would be. Meaning that, in this situation, the variables are maximally entangled and  imply maximal entanglement between pair members. If, on the other hand, we measure the momentum of the second emitted particle but not the first,  then  $P_{CM}$ would only be known  to accuracy $l_P/R_S^2$  since the kick from the first particle is unknown. Clearly, if we measure just the momentum of the $N^{\rm th}$ emitted particle (after, say, about half the particles were emitted so that $\;N \sim S_{BH}$), then  $P_{CM}$ would only be known  to an  accuracy of about $\; \sqrt{N}\frac{l_p}{R_S^2}\sim 1/R_S\;$. This would mean significant deviations of the pair from maximal entanglement.

What we then  need to consider is a superposition of bipartite states $\;|\vec P_{CM}(N),\vec q_N\rangle=|\vec P_{CM}(N)\rangle\;|\vec q_N\rangle\;$. Here, $\vec q_N$ is the momentum transfer in the $N^{{\rm th}}$ pair-production event
({\em i.e.}, the negative of  the net momentum of this produced pair)  and $\vec P_{CM}(N)$ is the center-of-mass momentum just before this event. Because of total momentum conservation, the latter is equal  to the sum of all previous momentum transfers, $\;\vec P_{CM}(N)=\sum\limits_i^{N-1} \vec q_i\;$.

It follows that we can write
\be
|\vec P_{CM}(N), \vec q_N\rangle\;=\;\frac{1}{\sqrt{\widetilde{\cal N}}}\;\int \int d^3 Q_N d^3 q_N \;\delta\left(\vec Q_N -\vec P_{CM}(N)-\vec q_N \right)\;  e^{-\frac{Q_N^2}{4\sigma_{CM}^2}}\;|\vec Q_N,\vec q_N\rangle \;,
\ee
where  $\;\vec Q_N = \sum\limits_i^N \vec q_i=\vec P_{CM}(N+1)\;$ and the  Gaussian incorporates the growing uncertainty in the center-of-mass position due to its  previously discussed random walk. As explained in Section~2, the width of this distribution is given by
\be
\sigma_{CM}^2(N) \;=\; N \sigma_i^2 \;\simeq\;  N \frac{l_p^2}{R_S^4} \;.
\label{sigma2}
\ee

A reduced density matrix for the momentum of the $N^{\rm th}$ pair can now
be written in the standard way,
\be
\rho_{red} \;=\; \int d^3 Q' \;\langle \vec Q'|\left[|\vec P_{CM}(N), \vec q_N\rangle
\langle\vec P_{CM}(N), \vec q_N|\right]| \vec Q'\rangle \;,
\ee
for which a straightforward calculation reveals that
\be
\rho_{red} \;= \; \frac{1}{{\cal N}}\int d^3 q_N \;
e^{-\frac{q_N^2}{2\sigma_{CM}^2}}\; |\vec q_N\rangle\langle \vec q_N|\;.
\label{reduced}
\ee

We may now apply Eq.~(\ref{reduced}) and the results of the Appendix, where it is shown that the degree of entanglement among the pair partners  is determined by the variance of the distribution $\sigma^2_{CM}$.
The purity of the reduced density matrix (in 3 space dimensions) is given by
$
{\rm Tr}\; \rho^2_{red}\;\simeq\; \left[\frac{\sigma_{CM}^2}{(\Delta Q)^2}\right]^{3/2}\;,
$
where $\Delta Q$ is the relevant range of momentum. Here, it is the Hawking temperature,  $\;\Delta Q\sim T_H\sim 1/R_S\;$, since the energies of the emitted particles are within the thermal window,
$
{\rm Tr}\; \rho^2_{red}\;\simeq\; \left[\sigma_{CM}^2 R_S^2\right]^{3/2}\;.
$

Using Eq.~(\ref{sigma2}), we obtain
\be
{\rm Tr}\; \rho^2_{red}\;\simeq\; \left[\frac{N}{S_{BH}(0)}\right]^{3/2}\;.
\label{purityII}
\ee
The purity is initially very small, indicating that the pairs are maximally entangled. As more particles are emitted, the purity increases until reaching  $\;N \sim S_{BH}(0)\;$. Then the purity becomes order one, indicating that the pairs are produced effectively in a product state.

These observations can be made more formal by parametrizing the associated entropy. Following Mathur \cite{MathurX},  one can define a parameter $\epsilon$ that indicates the deviation from maximal entanglement. In general, $\;\epsilon= \frac{S_{max}-S}{S_{max}}\;$, where $S$ is some measure of entropy and $S_{max}$ is its maximal value. Mathur used the Von Neumann entropy, whereas we used the R\'enyi entropy in \cite{schwing}. Different parametrizations will, of course, lead to somewhat different quantitative results.

In the current context, we find that the linear entropy $S_L$ is the appropriate measure of  entanglement, being directly related to the purity of the reduced density matrix
({\em e.g.}) \cite{linear}. For a system with a large number of possible states, as in the case of momentum entanglement, the linear entropy is given by  $\; S_L= 1-{\rm Tr}\;\rho^2_{red}\;$.   The  linear entropy for the  pairs is therefore given by
\be
S_L(N)\;\simeq\;
1-\left[\frac{N}{S_{BH}(0)}\right]^{3/2}\;.
\ee

As the linear entropy ranges from 1 (maximal entanglement) to 0 (product state),
the deviation parameter for the $N^{\rm th}$ produced pair is simply
\be
\epsilon(N)\;\simeq\; \left[\frac{N}{S_{BH}(0)}\right]^{3/2}\;.
\ee
As expected, this parameter is initially very small and already of order unity at times comparable to the Page time but never exceeds unity.

One final consideration:
As the entanglement between the pair partners decreases, the entanglement  between the positive-energy partner and the previously emitted particles will
increase in kind. This is because
the total momentum of all emitted particles must  sum up to zero by the end of evaporation; so that,
if we knew the momenta of the first $N$ emissions, our knowledge of the
$N^{\rm th}$ particle's momentum would accordingly increase.

This latter entanglement also measures the amount of correlation between the $N^{\rm th}$ emission and all the emissions that follow it until the end of the evaporation. This is because, if one measures the momenta of the first $N$ emitted articles, then the sum of momenta of the rest of the $S_{BH}(0)-N$ particles has to be exactly of the same magnitude and opposite direction as the sum of the first $N$ particles.

To make this idea precise, one can repeat the previous analysis for $\;\vec{K}_{CM}(N)=\sum\limits_{i=N+1}^{S_{BH}(0)} \vec{q}_i\;$. Then, as $N$ grows, the variance of $\vec{K}_{CM}(N)$ is decreasing  in proportion to the decrease of the variance in $\vec{P}_{CM}$.
It is also clear that $\vec{q}_{N}$ is always maximally entangled with the sum $\vec{P}_{CM}+\vec{K}_{CM}$ because $\;\vec{q}_N + \vec{P}_{CM}+\vec{K}_{CM}=0\;$.

\section{Conclusion}

We have  relied on a pair of  fundamental ideas ---  unitary evolution and
the fact that a finite mass BH must fluctuate quantum mechanically ---
to conclude  that the state of the near-horizon pairs is much different than the Unruh state at all but the earliest stages of BH evaporation. Mathur and others have deduced this outcome on the basis of  general arguments; our contribution is to provide a physical mechanism that leads to such a state and to provide a quantitative treatment of its deviations from the Unruh vacuum.
Our result is consistent with \cite{schwing}, where we found by using the
constraint of strong subadditivity  that $N/S_{BH}(0)$ is an  upper bound on the degree of disentanglement.

A central lesson of our work is that deviations from maximal entanglement depend on the quantum fluctuations of a finite-mass BH. On the contrary, an infinitely massive object cannot recoil, assuring that momentum  and, therefore, perfect entanglement are conserved for the produced pairs. This explains why Hawking and others concluded that the pairs were maximally entangled; given the  assumption of an infinitely massive BH, this must be so.

The techniques that were used here have an element of crudeness, as do many model-independent calculations.  One would like to further the analysis  in a more rigorous way, but this requires a much clearer  understanding  about the state of the BH interior and, then, how it changes when a negative-energy particle is subsumed. We have only begun to broach the subject of the BH interior \cite{inny} but hope that this path eventually leads to calculations along these lines.

\section*{Acknowledgments}

We would like to thank Samir Mathur for discussions.
The research of RB was supported by the Israel Science Foundation grant no. 239/10. The research of AJMM received support from an NRF Incentive Funding Grant 85353, an NRF Competitive Programme Grant 93595 and Rhodes Research Discretionary Grants. AJMM thanks Ben Gurion University for their  hospitality during his visit.

\appendix
\section{Momentum entanglement}

As a concrete example of momentum entanglement, we discuss the state of a pair of particles in 3 space dimensions. For simplicity, we consider a pair of massless bosons and, therefore, the state has to be symmetric under the exchange of the particles. Similar conclusions would be obtained for massive bosons or fermions.

The state of the pair can be written as
\be
|pair\rangle \;=\; 
 \int d^3 p_1 d^3 p_2 \  g(\vec{p}_1,\vec{p}_2) |\vec{p}_1,\vec{p}_2\rangle\;,
\ee
and its density matrix is then
\be
\widehat\rho_{pair}\;=\; 
 |pair\rangle\langle pair|\;,
\ee
so that the density-matrix elements are given by
\be
\rho(\vec{p}_1,\vec{p}_2;\vec{q}_1,\vec{q}_2)\;=\; g(\vec{p}_1,\vec{p}_2) g^\dagger(\vec{q}_2,\vec{q}_1)\;,
\ee
and the reduced density-matrix elements are expressible as
\be
\rho_{red}(\vec{p}_1,\vec{p}_2)\;=\;  (g g^\dagger)(\vec{p}_1,\vec{p}_2)\;.
\ee

Given that the full density matrix is normalized, the reduced density matrix is as well,  $\;{\rm Tr} \;g g^\dagger=1\;$.
One can use the  R\'enyi entropy  for estimating the amount of entanglement,
$\;H_2 =-\ln \frac{{\rm Tr} (\rho_{red}^2)}{\left({\rm Tr} \;\rho_{red}\right)^2}=-\ln {\rm Tr} \left[(g g^\dagger)^2\right]\;$.

If the total momentum of the pair vanishes, then
\be
g(\vec{p}_1,\vec{p}_2)\; =\; \frac{1}{\sqrt{{\cal N}}} \delta(\vec{p}_1+\vec{p}_2)\;,
\ee
where ${\cal N}$ is a normalization factor (${\rm Tr}\;1={\cal N}$).
The  R\'enyi entropy is then given by
\be
H_2 \;=\;  \ln {\cal N}\;,
\ee
which indicates that  the pair is  maximally entangled. As expected, when the total momentum of the pair is fixed, the state of the pairs is
indeed maximally entangled.
On the other hand, if the state is a product state,
\be
g(\vec{p}_1,\vec{p}_2) \;= \;\frac{1}{\sqrt{{\cal N}_1}}  f_1(\vec{p}_1)\frac{1}{\sqrt{{\cal N}_2}} f_2(\vec{p}_2)\;,
\label{prodstate}
\ee
with ${\cal N}_1$, ${\cal N}_2$ being normalization factors,
then $\;H_2=0\;$ as expected.

We now understand that, in order to deviate from maximal entanglement, there must be some spread in the total momentum of the pair.   In technical terms, the matrix $g(\vec{p}_1,\vec{p}_2)$ has to have some support away from the diagonal.
An example could be a Gaussian spread in the momentum difference with some small width $\sigma<\Delta Q$  where  $\Delta Q$ represents the window of applicable momenta,
\be
g(\vec{p}_1,\vec{p}_2)\;=\;\frac{1}{\sqrt{{\cal N}}} e^{-\frac{(\vec{p}_1-\vec{p}_2)^2}{2 \sigma^2}}\;.
\ee

In the case of this Gaussian, the purity of the reduced density is given by
\be
{\rm Tr}\; \rho^2_{red} \;=\; \frac{1}{{\cal N}}
 \int^{\Delta Q} d^3 p_1\; d^3 p_2\; d^3 p_3\; d^3 p_4 \;
e^{-\frac{(\vec{p}_1-\vec{p}_2)^2}{2 \sigma^2}}\; e^{-\frac{(\vec{p}_2-\vec{p}_3)^2}{2 \sigma^2}}\; e^{-\frac{(\vec{p}_3-\vec{p}_4)^2}{2 \sigma^2}}\; e^{-\frac{(\vec{p}_4-\vec{p}_1)^2}{2 \sigma^2}}\;,
\ee
where
$\;\sqrt{{\cal N}} ={\rm Tr}\;\rho_{red}=
 \int^{\Delta Q} d^3 p_1 d^3 p_2\;
e^{-\frac{(\vec{p}_1-\vec{p}_2)^2}{2 \sigma^2}}\;$
 and the upper limit on the integrals means  $\;\int^{\Delta Q} d^3 p =\int_{-\Delta Q}^{\Delta Q}
dp_x \int_{-\Delta Q}^{\Delta Q}
dp_y\int_{-\Delta Q}^{\Delta Q}
dp_z\;$.

For small values of the dimensionless variance,
$\;\frac{\sigma^2}{(\Delta Q)^2}<1\;$, one then obtains
\be
{\rm Tr}\;\rho^2_{red}\;=\; \left(\frac{\pi}{2}\right)^{3/2}
\left[\frac{\sigma^2}{(\Delta Q)^2}\right]^{3/2}
\left[1+ \frac{3(2-\sqrt{2})}{4\sqrt{\pi}}\sqrt{\frac{\sigma^2}{(\Delta Q)^2}}+\cdots\right]\;.
\ee

The R\'enyi entropy $(H_2) =-\ln({\rm Tr}\;\rho^2_{red})$ is then given by
\be
H_2 \;\simeq\; 3/2 \ln \left[\frac{(\Delta Q)^2}{\sigma^2}\right]\;.
\label{purity}
\ee
Here, it should be understood that this expression has to be normalized in the limiting case $\;\sigma^2\ll(\Delta Q)^2\;$.
We  circumvent the issue of normalization in the main text by using
the linear entropy to quantify the entanglement of the pairs.
 But what is clear is that
small values of the dimensionless variance
lead to small deviations from the  maximum entanglement.

If, on the other hand, the width $\sigma$ is large and extends over all
of  the allowed range of momenta,  then the state becomes effectively a product state as in Eq.~(\ref{prodstate}) with uniform $f_1$, $f_2$.
In this case, $H_2$ is small.
This outcome is already clear by taking the limit of $\;\sigma^2/(\Delta Q)^2\to 1\;$ in Eq.~(\ref{purity}), as then $\;{\rm Tr}\;\rho^2_{red}\to 1\;$ so that
$\;H_2\to 0\;$.


\begin{thebibliography}{99}



\bibitem{Hawk} S. W. Hawking, ``Black hole explosions'',  Nature {\bf 248}, 30  (1974); ``Particle creation
by black holes'',
Comm. Math. Phys. {\bf 43}, 199 (1975).




\bibitem{info}
  S.~W.~Hawking,
  ``Breakdown of Predictability in Gravitational Collapse,''
  Phys.\ Rev.\ D {\bf 14}, 2460 (1976).


\bibitem{Mald}
  J.~M.~Maldacena,
  ``Eternal black holes in anti-de Sitter,''
  JHEP {\bf 0304}, 021 (2003)
  [hep-th/0106112].



\bibitem{Pagerev}
  D.~N.~Page,
  ``Black hole information,''
  hep-th/9305040.


\bibitem{Mathur1}
  S.~D.~Mathur,
  ``What Exactly is the Information Paradox?,''
  Lect.\ Notes Phys.\  {\bf 769}, 3 (2009)
  [arXiv:0803.2030 [hep-th]].

\bibitem{Mathur2}
  S.~D.~Mathur,
``The Information paradox: A Pedagogical introduction,''
  Class.\ Quant.\ Grav.\  {\bf 26}, 224001 (2009)
  [arXiv:0909.1038 [hep-th]].


\bibitem{Mathur3}
S.  D.  Mathur,
 ``What the information paradox is {\it not},''
  arXiv:1108.0302 [hep-th].


\bibitem{Mathur4}
S. D. Mathur,
  ``What does strong subadditivity tell us about black holes?,''
  Nucl.\ Phys.\ Proc.\ Suppl.\  {\bf 251-252}, 16 (2014)
  [arXiv:1309.6583 [hep-th]].

\bibitem{MathurX}
 B.~D.~Chowdhury and S.~D.~Mathur,
  ``Pair creation in non-extremal fuzzball geometries,''
  Class.\ Quant.\ Grav.\  {\bf 25}, 225021 (2008)
  [arXiv:0806.2309 [hep-th]];
 B.~D.~Chowdhury and S.~D.~Mathur,
  ``Non-extremal fuzzballs and ergoregion emission,''
  Class.\ Quant.\ Grav.\  {\bf 26}, 035006 (2009)
  [arXiv:0810.2951 [hep-th]].



\bibitem{Mathur5}
  S.~D.~Mathur,
  ``Fuzzballs, Firewalls and all that...,'' to be published.




\bibitem{AMPS}
 A.~Almheiri, D.~Marolf, J.~Polchinski and J.~Sully,
  ``Black Holes: Complementarity or Firewalls?,''
  JHEP {\bf 1302}, 062 (2013)
  [arXiv:1207.3123 [hep-th]].




\bibitem{Sunny}
 N.~Itzhaki,
  ``Is the black hole complementarity principle really necessary?,''
 arXiv:hep-th/9607028.




\bibitem{Hooft}
  G.~'t Hooft,
  ``The Selfscreening Hawking atmosphere: A New approach to quantum black hole microstates,''
  Nucl.\ Phys.\ Proc.\ Suppl.\  {\bf 68}, 174 (1998)
  [gr-qc/9706058].



\bibitem{Braun}
 S.~L.~Braunstein, S.~Pirandola and K.~Zyczkowski,
  ``Entangled black holes as ciphers of hidden information,''
  Physical Review Letters 110, {\bf 101301} (2013)
  [arXiv:0907.1190 [quant-ph]].









\bibitem{MP}
D.~Marolf and J.~Polchinski,
  ``Gauge/Gravity Duality and the Black Hole Interior,''
  Phys.\ Rev.\ Lett.\  {\bf 111}, 171301 (2013)
  [arXiv:1307.4706 [hep-th]].



\bibitem{Bousso}
  R.~Bousso,
   ``Firewalls From Double Purity,''
  Phys.\ Rev.\ D {\bf 88}, 084035 (2013)
  [arXiv:1308.2665 [hep-th]];
 ``Frozen Vacuum,''
  Phys.\ Rev.\ Lett.\  {\bf 112}, 041102 (2014)
  [arXiv:1308.3697 [hep-th]].












\bibitem{page}
  D.~N.~Page,
  ``Average entropy of a subsystem,''
  Phys.\ Rev.\ Lett.\  {\bf 71}, 1291 (1993)
  [arXiv:gr-qc/9305007];
 ``Information in black hole radiation,''
  Phys.\ Rev.\ Lett.\  {\bf 71}, 3743 (1993)
  [arXiv:hep-th/9306083].





\bibitem{noburn}
 R.~Brustein and A.~J.~M.~Medved,
  ``Black hole firewalls, smoke and mirrors,''
  Phys.\ Rev.\ D {\bf 90}, no. 2, 024040 (2014)
  [arXiv:1401.1401 [hep-th]].




\bibitem{stick}
R.~Brustein and A.~J.~M.~Medved,
  ``Falling through the black hole horizon,''
  JHEP {\bf 1506}, 089 (2015)
  [arXiv:1503.05597 [hep-th]].




\bibitem{RB}
  R.~Brustein,
  ``Origin of the blackhole information paradox,''
  Fortsch.\ Phys.\  {\bf 62}, 255 (2014)
  [arXiv:1209.2686 [hep-th]].

\bibitem{gia}
  G.~Dvali and C.~Gomez,
  ``Black Hole's Quantum N-Portrait,''
  Fortsch.\ Phys.\  {\bf 61}, 742 (2013)
  [arXiv:1112.3359 [hep-th]].


\bibitem{Matt}
  M.~Visser,
  ``Physical observability of horizons,''
  Phys.\ Rev.\ D {\bf 90}, no. 12, 127502 (2014)
  [arXiv:1407.7295 [gr-qc]].




\bibitem{slowleak}
  R.~Brustein and A.~J.~M.~Medved,
  ``Restoring predictability in semiclassical gravitational collapse,''
  JHEP {\bf 1309}, 015 (2013)
  [arXiv:1305.3139 [hep-th]].

\bibitem{slowburn}
  R.~Brustein and A.~J.~M.~Medved,
  ``Phases of information release during black hole evaporation,''
  JHEP {\bf 1402}, 116 (2014)
  [arXiv:1310.5861 [hep-th]].


\bibitem{flameoff}
 R. Brustein and A. J. M. Medved,
``Horizons of Semiclassical Black holes are Cold,''
JHEP {\bf 1406}, 057 (2014) [arXiv:1312.0880[hep-th]].




\bibitem{dvali}
  G.~Dvali and C.~Gomez,
  ``Black Hole's Quantum N-Portrait,''
  Fortsch.\ Phys.\  {\bf 61}, 742 (2013)
  [arXiv:1112.3359 [hep-th]].



\bibitem{inny}
  R.~Brustein and A.~J.~M.~Medved,
  ``Quantum state of the black hole interior,''
  arXiv:1505.07131 [hep-th].


\bibitem{MathurXX}
  S.~D.~Mathur and D.~Turton,
  ``Oscillating supertubes and neutral rotating black hole microstates,''
  JHEP {\bf 1404}, 072 (2014)
  [arXiv:1310.1354 [hep-th]].









\bibitem{RM}
  R.~Brustein and M.~Hadad,
  ``Wave function of the quantum black hole,''
  Phys.\ Lett.\ B {\bf 718}, 653 (2012)
  [arXiv:1202.5273 [hep-th]].






\bibitem{flucyou}
 R.~Brustein and A.~J.~M.~Medved,
  ``Semiclassical black holes expose forbidden charges and censor divergent densities,''
  JHEP {\bf 1309}, 108 (2013)
  [arXiv:1302.6086 [hep-th]].

\bibitem{sunny}
N. itzhaki, private communication, 2009.

\bibitem{Mathurplusplus}
  E.~Keski-Vakkuri, G.~Lifschytz, S.~D.~Mathur and M.~E.~Ortiz,
  ``Breakdown of the semiclassical approximation at the black hole horizon,''
  Phys.\ Rev.\ D {\bf 51}, 1764 (1995)
  [hep-th/9408039].

\bibitem{Mathurplus}
 E.~Keski-Vakkuri and S.~D.~Mathur,
  ``Quantum gravity and turning points in the semiclassical approximation,''
  Phys.\ Rev.\ D {\bf 54}, 7391 (1996)
  [gr-qc/9604058].








\bibitem{PP1}
 P. Krekora Q. Su, and R. Grobe, ``Entanglement for pair production on
the zeptosecond scale,'' J. Mod. Optic. {\bf 52}, 489 (2005).



\bibitem{PP2}
M. V. Fedorov, M. A. Efremov and P. A. Volkov, ``Double and
multi-photon pair production and electron-positron entanglement,'' Opt.
Commun. {\bf 264}, 413 (2006).


\bibitem{linear}
T.-C. Wei, K.  Nemoto, P. M.  Goldbart, P. G. Kwiat,  W. J. Munro and
 F. Verstraete,
``Maximal entanglement versus entropy for mixed quantum states,''
Phys. Rev. A {\bf 67}, 022110 (2003)
[arXiv:quant-ph/0208138].

\bibitem{schwing}
R.~Brustein and A.~J.~M.~Medved,
  ``Constraints on the quantum state of pairs produced by semiclassical black holes,''
  JHEP {\bf 1507}, 012 (2015)
  [arXiv:1503.05351 [hep-th]].











\end{thebibliography}
\end{document}